\documentclass[aps,prb,twocolumn,amsfonts,amssymb,showpacs,eqsecnum]{revtex4}

\usepackage{amsmath}
\usepackage{amssymb}
\usepackage{graphicx}

\begin{document}

\title{External force affected escape of Brownian particles
from a potential well}
\author{A. I. Shushin}
\affiliation{Institute of Chemical Physics, Russian Academy of Sciences, 117977, GSP-1,
Kosygin str. 4, Moscow, Russia}

\begin{abstract}
The effect of an external force on the kinetics of
diffusion-assisted escaping of Brownian particles from a potential
well is analyzed in detail. The analysis is made within the
two-state model of the process which is known to be valid in the
deep well limit in the absence of external force. The generalized
variant of this model, taking into account the effect of the force,
is shown to be quite accurate as well for some shapes of the well.
Within the generalized two-state model simple expressions for the
well depopulation kinetics and, in particular, the for the escape
rate are obtained. These expressions show that the effect of the
force ($F$) manifests itself in the escape rate dependence on the
only parameter $\varphi = Fa/(2k_b T)$, where $a$ is the Onzager
radius of the attractive part of the well $U(r)$, defined by the
relation $|U(a)| = k_b T$.  The limiting behavior of this dependence
in the cases of weak and strong force is analyzed in detail.
Possible applications as well as the relation of the results of the
analysis to those obtained earlier are briefly discussed.
\end{abstract}

\pacs{82.20.Db, 82.20.Mj, 61.20.Lc} \maketitle

\bigskip

\section{Introduction}

The effect of external force on mechanisms and kinetic properties of
condensed phase diffusion-assisted reaction processes is considered
in a large number of works both experimentally and
theoretically.\cite{Ca,Ri} The active interest of scientists to this
phenomenon results from its great practical importance.

One of the most important systems, in which the external force
effect is investigated very thoroughly, is recombining geminate ion
pairs, undergoing relative diffusion in the external electric
field.\cite{Ri,Noo,Noo1} Most of theoretical studies analyze the
kinetics of the processes within the simplest model, which reduces
the problem to solving the Smoluchowski equation for probability
distribution function (PDF) of particles diffusing in a pure Coulomb
potential (with an external force) and reacting with the rate highly
localized at short distances.\cite{Ri,Street1} Even in this most
simple formulation the problem can, in general, be solved only
numerically, though detailed analytical analysis of some simple
variant of the problem have also been made,\cite{Ri,Street1} for
example, within the prescribed diffusion approximation.\cite{Mo}

Recent advances in time resolved investigations of charge transfer
and escaping processes in fast geminate reactions\cite{Chen} and, in
particular, geminate recombination of ion pairs in polar
liquids\cite{Bart1,Bart2,Shkrob,Chen,Crow,Ich,Peter} inspire further
development of theoretical methods of the analysis of the considered
problem. The main challenge of the theoretical studies consists in
the correct description of the manifestation of specific features of
the interparticle interaction (in real liquids) in the reaction
kinetics in a tractable form simple enough to be suitable for
applications.

In the majority of above mentioned theoretical works no specific
features of the form of the interaction potential for the probe
(Brownian) particles at short distances (of order of molecular size)
have been taken into account. In the condensed phase, however, the
distance dependence of the potential at short interparticle
distances $r$ can be strongly modified by interaction of particles
under study with those of the medium.\cite{Ri} This modified
interaction is usually characterized by the so called mean force
potential (MFP), which in a physically reasonable form incorporates
the medium effect and, in particular, discreteness of the medium at
short distances. The interaction with the medium particles is known
to result in the wavy behavior of the MFP at short distances.
Moreover, in some systems the medium effect results in the well-type
shape of the MFP at short distances (see Fig. 1) with the a markedly
high barrier at distances $r$ of order the distance of closest
approach $d$. This effect is found, for example, in the case ion
pairs in polar liquids.\cite{Ri,Nin}

Concerning the applicability of well-type approximation for the real
MFP shape, it is also worth mentioning the additional reasoning:
from mathematical and kinetic points of view any attractive
potential can be considered as well-shaped in the absence of (or
low) reactivity of particles at $r \sim d$. The only difference of
this type of wells from those shown in Fig. 1 is in their urge-like
shape at $r \sim d$.

\begin{figure}
\setlength{\unitlength}{1cm}
\includegraphics[height=5cm,width=7cm]{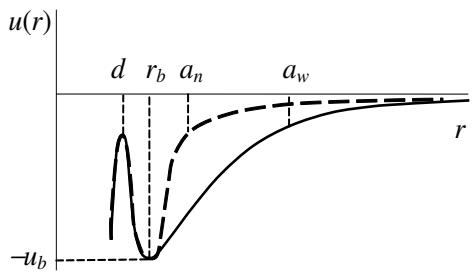}
\caption{The picture of the interaction potential $u(r)$ for two
models of its shape: narrow (dashed) and wide (full) well; $a_n$ and
$a_w$ are the Onzager radii for these models [defined by $u(a_{\nu})
= 1, \, (\nu = n, w)$], $d$ is the distances of closes approach, and
$r_{b}$ is the coordinate of the bottom.}
\end{figure}

The well-type shape of the MFP (with the reaction barrier at $r \sim
d$) results in the formation of the quasiequilibrium state within
the well, which can also be considered as a cage state. In the
absence of external force the kinetics of diffusion-assisted
depopulation of the initially populated cage state is analyzed in
detail in a number of papers.\cite{shu1,shu2,shu3,shu4} In the limit
of deep well depth the problem is shown to be accurately described
with the model of two kinetically coupled states: the
quasiequilibrium localized state within the well and the free
diffusion state outside the well.\cite{shu3}

The two-state model enables one to obtain the well depopulation
kinetics in a relatively simple analytical form. This kinetics,
determined by the monomolecular reactive passing over the barrier at
$r\sim d$ and escaping from well (cage), appears to be
non-exponential, in general. In the limit of deep well, however, the
deviation from the exponential kinetics is shown to be fairly
small.\cite{shu2,shu3}

In this paper we generalize the two-state model to describe the
effect of the external force on the well depopulation kinetics. The
effect is assumed to result only from the force induced change of
the rate of escaping from the well and will concentrate on the
discussion of the manifestation of the force in the escaping
kinetics.

Within the generalized two-state model we derive simple formulas for
the well depopulation (or escaping) kinetics for different models of
the well shape. Moreover, taking into account that in the limit of
deep well (which is of main interest of the discussion) the kinetics
is close to exponential, in our study we restrict ourself to the
analysis of the escape rate, only. The analysis shows that the
escape rate is strongly affected by the external force and the
specific manifestations of this effect depends on the well shape.
The force effect can be characterized by the rate dependence on the
only parameter. In the cases of weak and strong force the limiting
analytical expressions for this dependence are obtained and briefly
discussed.

\section{Formulation of the problem}

We consider the dynamics of the Brownian particle undergoing
diffusive motion in a three dimensional spherically symmetric MFP
well $U({\bf r}) \equiv U(r)$ centered at $r = 0$, where ${\bf r}$
is the vector of position of the particle. The dependence of $U(r)$
on the distance $r = |{\bf r}|$ is schematically shown in Fig.1. The
MFP will be characterized by three parameters: the distance $a$
(called hereafter the Onzager radius), at which $U(a) = k_B T$ [in
Fig. 1  $a_n$  and $a_w$ denote the Onzager radii, corresponding to
two models of the well shape, discussed below (in Sec. V)], the
distance of closest approach $d$, and the radius $r_b$ of the bottom
of the well whose energy is $U(r_b)= -U_b = -(k_B T) u_b$. At $r
\sim d$ the MFP $U(r)$ is assumed to be of the shape of a barrier,
diffusive crossing of which models the reaction within the well.

The main purpose of the work is to analyze the kinetics of
diffusion-assisted escaping from the well in the presence of the
external force $-{\bf F}$, i.e. escaping from the well of the
potential $U_{\bf f}=U(r)+({\bf F}\cdot{\bf r})$. For definiteness
the force is assumed to be directed along the axis $z$: ${\bf F}=
(0,0,F)$. The analysis can conveniently be made in spherical
coordinates in which ${\bf r}= (r\sin \theta \cos\phi, r \sin \theta
\sin\phi, r \cos\theta)$.

The diffusive space-time evolution of the Brownian particle is
described by the PDF $\rho ({\bf r},t)\equiv \rho
(r,\theta,\phi|t)$. In general, in spherical coordinates the PDF
depends on all three variables $r$, $\theta$, and $\phi$. However,
in the considered case of isotropic diffusion and isotropic initial
condition $\rho ({\bf r},t=0) = \rho_i (r)$ the PDF $\rho ({\bf
r},t)$ is independent of the azimuthal angle $\phi$ so that $\rho
({\bf r},t) \equiv \rho (r,\theta|t)$. In our analysis we will
assume that particles are created within the well at initial
distance $r_i \sim r_b$:
\begin{equation} \label{form0}
\rho ({\bf r},t=0)=(4\pi r_i^2)^{-1}\delta (r - r_i),
\end{equation}

The PDF $\rho (r,\theta|t)$ satisfies the Smoluchowski equation
\begin{equation} \label{form1}
\dot \rho =  \nabla_{\bf r}[ D (r) (\nabla_{\bf r} \rho +
\rho\nabla_{\bf r}u_{\bf f}) ],
\end{equation}
where $\nabla_{\bf r} $ is the gradient operator,
\begin{equation} \label{form2}
u_{\bf f}({\bf r}) =u(r)+({\bf f}\cdot{\bf r}),
\end{equation}
with $u(r)= U(r)/(k_B T)$ and ${\bf f}= {\bf F}/(k_B T)$, is the
dimensionless MFP, and $D (r)$ is the diffusion coefficient, which
in our analysis is assumed to be independent of $r$: $D(r) = D$
[though some possible effects of $D(r)$-dependence can also be
studied].

Note that the reaction kinetics for pairs of interacting Brownian
particles, say $a$ and $b$, is described by the equation similar to
eq. (\ref{form1}) with ${\bf r} = {\bf r}_a-{\bf r}_b$ and
parameters expressed in terms of those for separate
particles.\cite{Ri,Noo1}

In the absence of force the kinetics of escaping from the
spherically symmetric short range potential well $u (r)$ is analyzed
earlier.\cite{shu1,shu2,shu3,shu4} Here we extend the approach
applied in these works to describe the effect of external force.
This approach is based on the approximate solution of the eq.
(\ref{form1}) in the limit of deep well, in which the solution can
be obtained in analytical form by expansion in a small parameter
$\tau_r /\tau_e \ll 1$, where $\tau_r \sim (a-d)^2/D$ is the time of
equilibration within the well and $\tau_e \sim \tau_r e^{-u_a}$ is
the time of escaping from the well.\cite{shu2,shu3}

Analysis of this solution shows\cite{shu3} that in the lowest order
in the parameter $\tau_r /\tau_e$ the Smoluchowski approximation
(\ref{form1}) is equivalent to the model of two kinetically coupled
states: the state within the well and the state of free diffusion
outside the well. To clarify the main points of this model we will
first briefly consider the case of the absence of the force ${\bf F}
= 0$.

\section{Two-state model in the absence of force}

Originally, the two-state model was proposed to treat the kinetics
of the diffusion-assiisted escaping from the well in the case ${\bf
F} = 0$, in which the PDF $\rho ({\bf r},t)$ depends only the
distance $r = |{\bf r}|$: $\rho ({\bf r},t) \equiv \rho (r,t)$.

Analysis shows\cite{shu2,shu3} that in this case in the lowest order
in $\tau_r /\tau_e \ll 1$ the escaping kinetics can be described
within the model of two kinetically coupled states: highly localized
state within the well and free diffusion state outside the well. The
evolution of the state within the well ($d < r < a$) is determined
by the well population
\begin{equation} \label{nof2}
n(t) = 4\pi \int_d^{a} \! dr \,r^2 \rho (r,t),
\end{equation}
while the evolution of the state outside the well ($r > a$) is
governed by the distribution function $c(r,t)$. The functions $n(t)$
and $c(r,t)$ satisfy simple kinetic equations\cite{shu2,shu3}
\begin{subequations} \label{nof4}
\begin{eqnarray}
\dot n &=&S_lK_{+} c(a) - (K_- + w_r)n \qquad
 \label{nof4a}\\
\dot c &=& \hat L_r c + (S_{l}^{-1}K_{-} n - K_{+} c) \delta
(r-a),\qquad \label{nof4b}
\end{eqnarray}
\end{subequations}
which should be solved with the initial condition
\begin{equation} \label{nof8}
n(0) = 1 \quad \mbox{and} \quad c(r,0) = 0
\end{equation}
[implied by eq. (\ref{form0})] and  the boundary conditions for
$c(r,t)$ written as $\nabla_r c (r,t)|_{r=a} = 0$ (this condition
corresponds to reflection at $r = a$) and $c(r \rightarrow \infty) =
0.$ .

In eqs. (\ref{nof4}) $S_l = 4\pi a^2$ and
\begin{equation} \label{nof5}
\hat L_r = D r^{-2}\nabla_r(r^2 \nabla_r)
\end{equation}
is the radial part of the free diffusion operator, in which
$\nabla_r = \partial/\partial r$. The terms proportional to
$K_{\pm}$ describe the above-mentioned kinetic coupling
(transitions) between the state within the well, located at $r = a
$, and the free diffusion state outside the well. In the considered
limit $\tau_r /\tau_e \ll 1$ the transition rates $K_{\pm}$ satisfy
the relations:\cite{shu3}
\begin{equation} \label{nof6} K_{\pm}
\rightarrow \infty \quad \mbox{and} \quad K_{-}/K_{+} = K_e^0 =
a^2/Z_w,
\end{equation}
where
\begin{equation}\label{nof7}
Z_w = \int_{d<r<a}dr \, r^2 e^{-u(r)}
\end{equation}
is the partition function for the well $u(r)=U(r)/(k_B T)$. Formula
(\ref{nof6}), obtained by comparing equilibrium solutions of eqs.
(\ref{form1}) and (\ref{nof4}),\cite{shu3} represents the detailed
balance relation for transitions between the states of the two-state
model.

The term $w_r n (\epsilon)$ in eq. (\ref{nof4a}) describes the
effect of the first order reaction (in the well) with the rate
\begin{equation} \label{nof8a}
w_r = (D/Z_w) \left(\int_{r\sim d} \! dr \,
r^{-2}e^{u(r)}\right)^{-1} .
\end{equation}

Solution of eqs. (\ref{nof4}) yields\cite{shu2,shu3}
\begin{equation}\label{nof9}
n(t) = \frac{1}{2\pi i} \int\limits_{-i\infty+0}^{i\infty+0} \!\!
d\epsilon \frac{\exp (\epsilon t)}{\epsilon + w_r + a^2K_e
V(\epsilon)}.
\end{equation}
The function $V(\epsilon)$ is directly related to the Green's
function of the operator controlling diffusion outside the well
[with the reflective boundary condition $\nabla_r g|_{r=a} =
0$]\cite{shu2,shu3}
\begin{equation} \label{nof10}
G(r,r_i|\epsilon) = \langle r | (\epsilon - \hat L_r)^{- 1} |r_i
\rangle :
\end{equation}
\begin{equation} \label{nof11}
V(\epsilon) = 1/G(a,a|\epsilon) = D [a^{-1} + (\epsilon/D)^{1/2}].
\end{equation}

Substitution of formula (\ref{nof11}) into eq. (\ref{nof9}) leads to
the following formula for the well population $n(t)$:\cite{shu3}
\begin{equation}
n(t)= \frac{1}{2\pi i} \int_{-i\infty+0}^{i\infty+0} d\varepsilon
\frac{\exp[\varepsilon (w_{0} t)]}{1 + \varepsilon + \gamma
\varepsilon^{1/2} }\label{nof12}
\end{equation}
where \, $\gamma = (w_e/w_{0})^{1/2}(a^2w_e/D)^{1/2}$. The rate
\begin{equation} \label{nof14}
w_{0} = w_e^0 + w_r
\end{equation}
is a sum of the rate of escaping from the well
\begin{equation} \label{nof15}
w_e^{0} =  \frac{D}{Z_w} \left(\int_{r_b}^{\infty} \! dr \,
r^{-2}e^{u(r)}\right)^{-1} = Da/Z_w
\end{equation}
and the rate of reaction in the well $w_r$ [see eq.(\ref{nof8a})].

The function $n(t)$ (\ref{nof12}) is, evidently, non-exponential.
Its analytical properties are essentially determined by the
parameter $\gamma \approx a/l_D$, where $l_D = (D/w_0)^{1/2}$ is the
average distance of diffusive motion during the lifetime $\tau_0 =
w_0^{-1}$ of the particle in the well. It is easily seen that for
deep wells $l_D \gg a$, and therefore $\gamma \ll 1$. The parameter
$\gamma$ controls the onset of the change of the functional form of
$n(t)$: $n(t) = \exp (-w_{0} t)\:\mbox{at} \: \tau \lesssim \ln
(1/\gamma)$ and $n(t) \sim 1/t^{3/2} \: \mbox{at} \: \tau \gg \ln
(1/\gamma)$.

It is worth emphasizing that {\it despite the complexity of the
escaping kinetics $n(t)$, in general, it is quite close to the
exponential for deep wells (for $\gamma \ll 1$), as it follows from
the above expressions}. Moreover, independently of the well depth
average steady state characteristics of type of the time averaged
kinetic parameters just coincide with those corresponding to the
exponential part of the kinetics. For example, the lifetime of
particles in the well
\begin{equation} \label{nof16}
\tau_0 = \int_{0}^{\infty}\!dt\, n(t) = w_0^{-1}.
\end{equation}

This means, in particular, that the steady state characteristics are
quite informative and can be used in the discussion of the escaping
kinetics in the more complicated case of the presence of an external
forces.

\section{Two-state model in the presence of force}

\subsection{Kinetic equations}

In the presence of force [${\bf f} = {\bf F}/(k_B T)\neq 0$] the
potential $u_{\bf f}(r)=u(r)+({\bf f}\cdot{\bf r})$ in eq.
(\ref{form1}), is not isotropic, which results in the dependence of
the solution $\rho ({\bf r},t)$ on particle polar angle $\theta$:
$\rho ({\bf r},t) \equiv \rho (r,\theta|t)$.

It is important to note that in the case ${\bf f} \neq 0$ the
two-state model is also valid for a variety of shapes of the
potential well $u (r)$ though some additional analysis of the
corresponding validity criteria are certainly required.

Similarly to the case of the absence of force, in any variant of the
anisotropic two-state model (for ${\bf f} \neq 0$) the kinetics is
also described by two functions: the population $n(\theta|t)$ within
the well and the PDF $c(r,\theta|t)$ of particles outside the well.
These functions, however, depend on the polar angle $\theta$. Within
the two-state model the effect of the force ${\bf f}$ shows itself
in the dependence of kinetic parameters on $\theta$. The form of
this dependence is determined by the particular variant of the
model.

In general, two-state kinetic equations, describing evolution of
PDFs $n(\theta|t)$ and $c(r,\theta|t)$ in the presence of an
external force, can be written by analogy with eqs. (\ref{nof4}):
\begin{subequations} \label{prf7}
\begin{eqnarray}
\dot n &=& S_l K_{+}(\theta) c(a|t) + [\hat L_{c}-( K_{-}(\theta) +
w_{r})]n, \qquad
 \label{prf7a}\\
\dot c &=& \hat L_{f} c + [S_{l}^{-1}K_{-}(\theta) n - K_{+}(\theta)
c] \delta (r-a),\qquad \label{prf7b}
\end{eqnarray}
\end{subequations}
where
\begin{equation} \label{prf8}
\hat L_{f} = D\nabla_{\bf r}(\nabla_{\bf r}  + {\bf f})
\end{equation}
is the operator, describing diffusion outside the well, and $\hat
L_c$ is the Smoluchowski operator in $\{\theta\}$-space which
controls orientational relaxation of the PDF in the well.

The essential difference of anisotropic equations from isotropic
ones consists in the orientation dependence of rates,
$K_{+}(\theta)$ and $K_{-}(\theta)$. Below, in accordance with eq.
(\ref{nof6}), in the considered limit $\tau_r /\tau_e \ll 1$ we will
assume the transition rates $K_{\pm}$ to satisfy the
relations:\cite{shu3}
\begin{equation} \label{nof7c}
K_{\pm} \rightarrow \infty \;\;\mbox{and}\;\;
K_{-}(\theta)/K_{+}(\theta) = K_e(\theta).
\end{equation}
Therefore in this limit $\theta$-dependence of rates
$K_{\pm}(\theta)$ show itself in that of the the equilibrium
constant: $K_e(\theta)$. The form of the function $K_e(\theta)$ is
determined by the shape of the well. Some model well shapes and
corresponding $K_e(\theta)$ dependences, as well as applicability of
the corresponding two-state models, are discussed below.

Equations (\ref{prf7}) should be solved with boundary conditions
\begin{equation} \label{prf9}
(\nabla_r + f\cos\theta) c|_{r=a}=0
\;\;\mbox{and}\;\;c|_{r\to\infty} \to 0,
\end{equation}
first of which describes reflection of particles (diffusing in the
state outside the well) at $r = a$. The initial condition is assumed
to be isotropic and given by eq. (\ref{form0}).

In what follows it will be convenient to represent functions
$n(\theta|t)$ and $c(r,\theta|t)$ in the form of vectors $|{\bf n}
(t)\rangle$ and $|{\bf c} (r,t)\rangle$, whose components are
obtained by expansion of these functions in the orthonornmal basis
of properly normalized Legendre polynomials (spherical
functions)\cite{Lan}
\begin{subequations} \label{prf1}
\begin{eqnarray} 
|Y_l\rangle &=& (l + 1/2)P_l (\cos\theta), \label{prf1a}\\
 \langle Y_l| &=& \int_0^{\pi}\!d\theta \, \sin\theta P_l
 (\cos\theta) \dots \label{prf1b}
\end{eqnarray}
\end{subequations}
with $(l=0,1,\dots):$
\begin{equation} \label{prf2}
|{\bf n}\rangle = \sum_{l=0}^{\infty} n_l |Y_l\rangle  \;\;
\mbox{and} \;\; |{\bf c}\rangle  = \sum_{l=0}^{\infty}  c_l
|Y_l\rangle,
\end{equation}
where for any vector $|\chi (\theta)\rangle,\; (\chi = n, c),$ its
components $\chi_l$ are defined by
\begin{equation} \label{prf3}
\chi_l  = \langle Y_l | \chi\rangle = \int_{0}^{\pi} \!\! d\theta\,
\sin\theta\, P_l(\cos\theta)\chi(\theta).
\end{equation}

In term of this vector representation the initial condition can
conveniently be written in the form, explicitly displaying its
independence of orientation:
\begin{equation} \label{prf10}
|\rho_i \rangle =(2\pi r_i^2)^{-1}|Y_0\rangle \delta (r - r_i).
\end{equation}

As for the initial condition, it is worth noting, in addition, that
in the most realistic limit of orientational relaxation within the
well much faster than the escaping from the well the escaping
kinetics is insensitive to the orientational dependence of the
initial condition.

\subsection{Escaping kinetics}

Equations (\ref{prf7}) can be solved by the method, applied above in
the case of the absence of force, but with the use of expansion of
$n(\theta|t)$ and $c(r,\theta|t)$ in spherical functions
$|Y_l\rangle$, i.e. vector representation $|{\bf n}(t)\rangle$ and
$|{\bf c}(r|t)\rangle$ [see eq. (\ref{prf2})]. The solution yields
for the Laplace transform
\begin{equation} \label{narw1}
|\widetilde{\bf n}(\epsilon)\rangle =  \int_0^{\infty}\!dt
e^{-\epsilon t}|{\bf n}(t)\rangle:
\end{equation}
\begin{equation} \label{narw2}
|\widetilde{\bf n}(\epsilon)\rangle =  \big(\epsilon + w_r + \hat
L_{c} + a^2\hat V(\epsilon)\hat K_e\big)^{-1}|{\bf n}_i \rangle,
\end{equation}
where for the initial condition (\ref{prf10}) $|{\bf n}_i \rangle =
(1/2\pi) |Y_0\rangle $. In this expression  the equilibrium constant
$\hat K_e$ is the operator, which indicates its dependence on the
orientation angle $\theta$, and
\begin{equation} \label{narw4}
\hat V =\hat G^{-1}(a,a|\epsilon)
\end{equation}
is the operator in the space of $|Y_l\rangle$-states, expressed in
terms of the evolution operator for diffusive motion outside the
well (evaluated at $r = r_i = a$):
\begin{eqnarray} \label{narw5}
\hat G(a,a|\epsilon) &=& \langle a | (\epsilon - \hat
L_f)^{- 1} |a \rangle\nonumber\\
&=& e^{-\varphi\cos\theta}\langle a | (\epsilon - \hat \Lambda_f)^{-
1} |a \rangle e^{\varphi\cos\theta},\qquad
\end{eqnarray}
in which $\varphi = fa/2$ and
\begin{equation} \label{narw6}
\hat \Lambda_f = D\big(\hat L_r + r^{-2}\hat
L_{\theta}-\mbox{$\frac{1}{4}$}f^2\big),
\end{equation}
with
\begin{equation} \label{narw7}
\hat L_{\theta} =
(1/\sin\theta)\nabla_{\theta}[\sin\theta(\nabla_{\theta})].
\end{equation}

In the expression (\ref{narw7}) $\hat L_r$ is the operator of radial
diffusion, defined in eq. (\ref{nof5}), and $\hat L_{\theta}$ is the
operator of free orientational diffusion, diagonal in
$|Y_l\rangle$-basis:
\begin{equation} \label{narw8}
\hat L_{\theta} = -\sum\nolimits_{l=0}^{\infty}l(l+1)|Y_l \rangle
\langle Y_l|.
\end{equation}

In what follows we will mainly restrict ourselves to the analysis of
the total population of the well $n_0(t)$, for the Laplace transform
of which one gets the expression
\begin{eqnarray} \label{narw8a}
\widetilde{n}_0(\epsilon) &=& 2\pi\!\int_0^{\pi}\!\!d\theta\,
\sin\theta \widetilde{n}(\theta,\epsilon) \equiv
2\pi\langle Y_0|\widetilde{\bf n}(\epsilon)\rangle\nonumber\\
& = &\langle Y_0|\big[\epsilon + w_r + \hat L_{c} + a^2\hat
V(\epsilon) \hat K_e\big]^{-1}|Y_0 \rangle,\;\;\quad
\end{eqnarray}

Moreover, since the escaping kinetics appears to be fairly close to
the exponential (see Sec. III), to characterize the process it is
sufficient to calculate the mean inverse lifetime $\bar{\tau}_0$ of
particles in the well defined by eq. (\ref{nof16}):
\begin{equation} \label{narw8b}
\bar{\tau}_0  = \widetilde{n}_0(0) = \langle Y_0|\big[w_r + \hat
L_{c} + a^2\hat V(0) \hat K_e\big]^{-1}|Y_0 \rangle.
\end{equation}

\subsection{General formulas}

Formulas (\ref{narw5}-(\ref{narw7}) allow us to evaluate the
operator $\hat V(\epsilon)$ in analytical form and, therefore,
analyze the behavior of $\widetilde{n}_0(\epsilon)$ and
$\bar{\tau}_0$ relatively easily.

In the evaluation it is worth taking into account the useful
relation which simplifies the differential operator in the radial
space:
\begin{equation} \label{narw9}
\langle a | (\epsilon - \hat \Lambda_f)^{- 1} |a \rangle = \langle a
| (\epsilon - \hat \lambda_f)^{- 1} |a \rangle,
\end{equation}
where
\begin{equation} \label{narw11}
\hat \lambda_f = D(\nabla_r^2 + r^{-2}\hat
L_{\theta}-\mbox{$\frac{1}{4}$}f^2).
\end{equation}

The evolution operator $\langle a | (\epsilon - \hat \lambda_f)^{-
1} |a \rangle$ can be obtained in analytical form\cite{shu5} with
the use of two linearly independent operator solutions $\hat
\psi_{-} (r)$ and $\hat \psi_{+} (r)$ of equation
\begin{equation} \label{narw12}
(\epsilon - \hat \lambda_l)\hat {\bf \psi}_{\pm} = 0
\end{equation}
in which the operator $\hat L_{\theta}$ is treated as a parameter.
These solutions satisfy two boundary conditions corresponding to
those given in eq. (\ref{prf9}) [after change of variable $c(r) =
e^{-(fr\cos\theta)/2}\,\psi (r)$]
\begin{equation} \label{narw14}
(\nabla_r + \mbox{$\frac{1}{2}$}f\hat \omega) \hat \psi_{-}|_{r=a}=0
\;\;\mbox{and}\;\; \hat \psi_{+}|_{r\to\infty} \to 0,
\end{equation}
where
\begin{equation} \label{narw15}
\hat \omega = \sum_{l,m=0}^{\infty} |Y_l\rangle \langle Y_l
|\cos\theta |Y_m\rangle\langle Y_m |
\end{equation}
is the matrix representation of the function $\cos\theta$. The
matrix elements $\langle Y_l|\cos\theta |Y_m\rangle = \langle
Y_l|P_1(\cos\theta) |Y_m\rangle$ are evaluated
analytically\cite{Lan} though the corresponding formulas will not be
needed in our further analysis.

Both solutions $\hat \psi_{+} (r)$ and $\hat \psi_{-} (r)$ are
expressed in terms of matrices of modified Bessel functions\cite{Abr} 
\begin{eqnarray} 
\hat K (r) &=& \sqrt{r}\sum\nolimits_{l=0}^{\infty}|Y_l \rangle
K_{l+\frac{1}{2}}(kr)\langle Y_l|,\label{narw17a}\\
\hat I (r) &=& \sqrt{r}\sum\nolimits_{l=0}^{\infty}|Y_l \rangle\,
I_{l+\frac{1}{2}}(kr)\,\langle Y_l|: \label{narw17b}
\end{eqnarray}
\begin{equation} \label{narw18}
\hat \psi_{+} (r) = \hat K (r),\;\;\hat \psi_{-} (r) = \hat I (r) +
\hat K (r)\hat\kappa,
\end{equation}
where
\begin{equation} \label{narw18a}
k(\epsilon) = \sqrt{(f/2)^2 +\epsilon/D}
\end{equation}
and $\hat \kappa$ is the matrix determined by the boundary condition
at $r =a$ [see eq. (\ref{narw14}]:
\begin{equation} \label{narw19}
\hat\kappa = [\nabla_r \hat K(r) -\hat q \hat K(r)]^{-1}[\hat q \hat
I(r)-\nabla_r \hat I(r)]|_{r=a},
\end{equation}
in which
\begin{equation} \label{narw20}
\hat q = a^{-1}(1 - \varphi \hat \omega)\equiv a^{-1}(1 - \varphi
\cos\theta),
\end{equation}
with $\varphi = fa/2$.

It is worth noting that the matrices $\hat K$ and $\hat I$ do not
commute with $\hat \omega$ and, therefore, the order of matrices in
the products of the matrices in expressions
(\ref{narw18})-(\ref{narw19}) is important. As a result of these
special commutation properties of the matrices, the matrix solutions
$\hat \psi_{+} (r)$ and $\hat \psi_{-} (r)$ do not commute either.

The general representation of the evolution operator  $\langle r |
(\epsilon - \hat \lambda_f)^{- 1} |r_i \rangle$ in terms of
non-commuting solutions $\hat \psi_{+} (r)$ and $\hat \psi_{-} (r)$
is proposed and thoroughly discussed in ref. [20]. This
representation generalizes the one well known in the case of scalar
solutions $\psi_{+} (r)$ and $\psi_{-} (r)$. In general, the
proposed representation is fairly complicated, however, in the
particular case of solutions given by eqs. (\ref{narw17a}) and
(\ref{narw17b}) it reduces to a more simple one:
\begin{widetext}
\begin{equation} \label{narw21}
\langle r | (\epsilon - \hat \lambda_f)^{- 1} |r_i \rangle = -
D^{-1}[\hat K (r)\hat I(r_i)\theta_H(r-r_i) + \hat I(r)\hat K
(r_i)\theta_H(r_i-r)+\hat K (r)\hat \kappa \hat K(r_i)]/W(\hat K,
\hat I).
\end{equation}
\end{widetext}
In this expression $\theta_H(x)$ is the Heaviside step function and
\begin{equation} \label{narw22}
W(\hat K, \hat I) = \nabla_r \hat K(r) \hat I(r) - \nabla_r \hat
I(r) \hat K(r)
\end{equation}
is the Wronskian of two solutions which is a scalar function
[$W(\hat K, \hat I) = -1$]. The validity of the expression
(\ref{narw21}) can easily be verified by direct substitution of it
to equation similar to eq. (\ref{narw12}) but with delta-function in
the right hand side.

For the particular case $r=r_i=a$ formula (\ref{narw21}) yields
\begin{equation} \label{narw23}
\langle a | (\epsilon - \hat \lambda_f)^{- 1} |a \rangle =
D^{-1}[\hat q - \hat q_K(\epsilon)]^{-1},
\end{equation}
where
\begin{equation} \label{narw24}
\hat q_K(\epsilon) = \nabla_r \hat K(r)/\hat K(r)|_{r=a} =
\sum_{l=0}^{\infty}|Y_l\rangle q_{K_l}(\epsilon) \langle Y_l|
\end{equation}
with
\begin{equation} \label{narw25}
q_{K_l}(\epsilon) = (l/a) + k(\epsilon) K_{l-\frac{1}{2}}
(k(\epsilon)a)/K_{l+\frac{1}{2}} (k(\epsilon)a).
\end{equation}

Substituting the expression (\ref{narw23}) into eq. (\ref{narw9})
and then into eqs.(\ref{narw6}) and (\ref{narw4}) we obtain fairly
simple formula for $\hat V(\epsilon)$:
\begin{eqnarray} \label{narw26}
\hat V(\epsilon) &=& D e^{-\varphi\cos\theta}(\hat q - \hat q_K)
e^{\varphi\cos\theta}\nonumber\\
&=& D[a^{-1}(1 - \varphi \cos\theta) - e^{-\varphi\cos\theta}\hat
q_K e^{\varphi\cos\theta}] .\qquad
\end{eqnarray}

For out further analysis of the escaping kinetics $n_0 (t)$ we need
to specify of the operator $\hat L_c$ describing orientational
relaxation in the well. Naturally it should be of the
Smoluchowski-like form:
\begin{equation} \label{narw27}
\hat L_{c} = D_c(\sin\theta)^{-1}
\nabla_{\theta}[\sin\theta(\nabla_{\theta} + \nabla_{\theta}\bar
u)],
\end{equation}
where $D_c$ is the orientational diffusion coefficient $\bar u
(\theta)$ is the effective orientational which is determined by the
shape of the well (see below).

For the sake generality, we will not assume any relation of the
value of $D_c$ with that of the diffusion coefficient $D$ outside
the well (of type of $D_c \sim D/r_b^2$).

Moreover, in the considered limit of large well depth it is quite
natural to assume that orientational relaxation is much faster than
well depopulation.

\subsection{Fast orientational relaxation in the well}

The fast orientational relaxation limit implies that $D_c \gg w_0$,
where $w_0$ is the rate of depopulation of the well defined in eq.
(\ref{nof16}). This relation means that after some time $\sim \tau_c
=D_c^{-1}$ of orientational relaxation (of the initial population in
the well) the vector of well population $|{\bf n}(t)$ remains close
to the equilibrium one $|\Psi_e\rangle$ during the process:
\begin{equation} \label{narw28}
|{\bf n}(t)\rangle \approx n_0(t) |\Psi_e\rangle,
\end{equation}
where
\begin{equation} \label{narw29}
|\Psi_e\rangle = Z_{\theta}^{-1}e^{-\bar{u}_b(\theta)},
\;\;\;\; Z_{\theta} =\int_{0}^{\pi}\!\!d\theta \,\sin\theta
e^{-\bar{u}_b(\theta)}.
\end{equation}
Note that  within bra-ket notation the adjoint vector $\langle
\psi_e |$ coincides with $\langle Y_0 |$ and is given by formula
\begin{equation} \label{narw29a}
\langle \Psi_e | = \langle Y_0 | = \int_0^{\pi}\!d\theta \,
\sin\theta \dots,
\end{equation}
which can be confirmed by the relation $\langle \Psi_e |\hat L_{c} =
0$ directly following from the definition of $\hat L_{c}$ [see eq.
(\ref{narw27})]. With the use of this formula one can easily find
that $|\psi_e \rangle$ satisfies the normalization condition
$\langle \Psi_e|\Psi_e \rangle = 1$.

In what follows we will restrict ourselves to the analysis of the
escaping kinetics just in this limit of fast orientational
relaxation.

For fast orientational relaxation the splitting $\delta L_c$ of
eigenvalues of the operator $\hat L_c$ ($\delta L_c \sim D_c$) is
much larger than $a^2 \| \hat V \hat K_e \| \sim w_e$. In such a
case in the lowest order in the parameter $w_e/D_c \ll 1$ we can
significantly simplify the general expressions (\ref{narw8a}) and
(\ref{narw8b}) for $\widetilde{n}_0 (\epsilon)$ and the inverse
average lifetime ${\bar w}_0 = {\bar \tau}_0^{-1} =
\widetilde{n}_0^{-1} (\epsilon)$, respectively, as follows:
\begin{equation} \label{narw30}
\widetilde{n}_0(\epsilon)  = \big[\epsilon + w_r  + a^2\langle
\Psi_e|\hat V(\epsilon) \hat K_e|\Psi_e\rangle]^{-1},
\end{equation}
and
\begin{equation} \label{narw30a}
{\bar w}_0 = {\bar \tau}_0^{-1}   = w_r  + a^2\langle \Psi_e|\hat
V(0) \hat K_e|\Psi_e\rangle,
\end{equation}

Equation (\ref{narw30}) presents the main result of the work for the
kinetics of the well depopulation in the limit of fast orientational
relaxation.

Formula (\ref{narw30}) shows that, in general, in the fast
orientational relaxation limit the escaping kinetics $n_0(t)$ is
non-exponential, however, as it has been mentioned above in Sec. III
in the considered case of deep wells its deviation from exponential
is very small and, therefore, the kinetics can quite accurately be
characterized by the only parameter, the mean lifetime ${\bar
\tau}_0^{-1} $ or the corresponding mean rate ${\bar w}_0 = {\bar
\tau}_0^{-1} $. For this reason, in what follows we will mainly
discuss these two parameters.

Moreover, below we will concentrate on the second term in square
brackets in eq.(\ref{narw30a}):
\begin{equation} \label{narw30b}
w_e = a^2\langle \Psi_e|\hat V(0) \hat K_e|\Psi_e\rangle,
\end{equation}
which can be interpreted as the mean escape rate.

To apply formulas (\ref{narw30})-(\ref{narw30b}) one need to specify
the shape of the potential well which determines the orientational
potential $\bar u (\theta)$ in the Smoluchowski-type operator $\hat
L_c$ [see eq. (\ref{narw27})] and, therefore, the equilibrium state
$|\Psi_e\rangle $. In our work we will consider two realistic models
of the well shape in which relatively simple analytical expressions
for the mean rate ${\bar w}_0 = {\bar \tau}_0^{-1} $ and, in
particular, for $w_e$ can be obtained.

\section{Application of general results}

In this section we will analyze the specific features of mean escape
rate $w_e$ for two variants of the well shape:

1) Narrow well shape (shown in Fig. 1 by dashed line with $a =
a_n$), for which $a - d \ll d$ and the time of equilibration within
the well $\tau_r \sim (a-d)^2/D \ll a^2/D \ll \tau_e$, where $\tau_e
\sim {\bar \tau}_0^{-1} $ [see eq. (\ref{nof16})] is the time of
escaping from the well;

2) The wide well shape (full line in Fig. 1 with $a = a_w$),
corresponding to a small distance of closest approach $d \ll a$ (in
which of the main interest is the region  $fd \ll 1$ while $fa
\gtrsim 1$). This shape is schematically shown in Fig. 1 by full
line with $a = a_w$.

The analysis will be made in the above-discussed limit of fast
orientational relaxation in the well with the use of eq.
(\ref{narw30}) for the well depopulation kinetics. In our study, in
addition to this formula and other ones derived above, we will also
use the representation for $ e^{\pm\varphi \cos\theta}$ in terms of
expansion in spherical functions $P_{l}(\cos\theta)$:\cite{Abr}
\begin{equation} \label{narw31}
 e^{\pm\varphi\cos\theta} = \! \sqrt{\frac{2\pi}{\varphi}}\,\sum_{l=0}^{\infty}(\pm
 1)^{l} (l + \mbox{$\frac{1}{2}$}) I_{l+\frac{1}{2}}(\varphi)P_{l}(\cos\theta).
\end{equation}

\subsection{Narrow-well shape}

In the case of narrow well, when $a -d < d$, the well is of the
shape of attractive well layer near the distance of closest approach
$d$. In this limit within the wide region force strengths $f <
1/(a-d)$ we can neglect the effect of the force on the radial shape
of the well and take into consideration only the dependence of well
depth $\bar{u}_(\theta)$ on the orientation angle $\theta$:
\begin{equation} \label{narw32}
{\bar u}_b(\theta) \approx u_{\bf f}(r_b,\theta) \approx
u_b+fa\cos\theta
\end{equation}
with $f = |{\bf f}| > 0$, and the force effect on free diffusion in
the state outside the well. In eq. (\ref{narw31}) we took into
account the smallness of the width of the well, $a - d \ll d$, which
leads to the high accuracy of the relation $fr_b \approx fa$.

It is important to note that the small value of the well width and,
therefore, fast equilibration of the well population in radial
direction, ensures the validity of the description of the kinetics
in terms of the angular coordinate dependent well population
$n(\theta|t)$ introduced above. Noteworthy is also that the
negligible force affected change of the well shape results in the
absence of the dependence of the detailed balance relation and the
equilibrium constant [see eq. (\ref{nof6})] on the angle $\theta$,
i.e $K_e =K_e^0$ . In other words, the effective partition function
$Z_w$, which governs the constant $K_e$, is still given by eq.
(\ref{nof7}), i.e. is controlled by the shape of the potential
$u(r)$ without external force, despite possible strong force effect
on the energy of the bottom predicted by eq. (\ref{narw32}). This is
because the external force leads to the identical change of both the
bottom energy ${\bar u}_b(\theta) \approx u_b+fa\cos\theta$ and the
energy of the free diffusion state at $r = a$: $u_{\bf f}(a,\theta)
\approx fa\cos\theta$.

The potential ${\bar u}_b(\theta)$ determines the kinetics of
orientational relaxation of the population in the well, which is
described by the Smoluchowski operator (\ref{narw27}) with
\begin{equation} \label{narw33}
{\bar u}(\theta)={\bar u}_b(\theta) - u_b=2\varphi\cos\theta,
\;\;\mbox{where}\;\; \varphi = fa/2.
\end{equation}
In this case the equilibrium state within the well is written as
\begin{equation} \label{narw34}
|\Psi_e\rangle = Z_{0}^{-1}(\varphi)e^{-2\varphi\cos\theta}
\end{equation}
with
\begin{equation} \label{narw34a}
Z_{0} (\varphi) = \sinh (2\varphi)/\varphi.
\end{equation}

Substitution of formulas (\ref{narw34}) and (\ref{narw31}) into the
expression (\ref{narw30b}) yields for the mean escape rate $w_e
(\varphi)$, expressed in terms of the universal function
$Q_n(\varphi)$:
\begin{equation} \label{narw35}
w_e (\varphi)/w_{e_n}^0 = Q_n(\varphi) = 
\mbox{$\frac{1}{2}$}+ \varphi\coth(2\varphi) + S_n(\varphi).
\end{equation}
In this formula  $w_{e_n}^0 = w_e^0$ is the escape rate in the
absence of a force [see eq. (\ref{nof15})], $\varphi =fa/2$, and
\begin{equation} \label{narw36}
S_n(\varphi)= 2Z_{0}^{-1}(\varphi)(\pi/\varphi)
\sum_{l=0}^{\infty}(l+\mbox{$\frac{1}{2}$})
I_{l+\frac{1}{2}}^2(\varphi) q_l(\varphi)
\end{equation}
with $Z_{0}(\varphi)$ defined in eq. (\ref{narw34}) and
\begin{equation} \label{narw39}
q_l (\varphi) = aq_{K_l} = l+\varphi K_{l-\frac{1}{2}}
(\varphi)/K_{l+\frac{1}{2}}(\varphi).
\end{equation}

Formula (\ref{narw35}) shows that the force effect on the rate is
characterized by the only parameter $\varphi = fa/2$. The numerical
calculated universal function $Q_n(\varphi)$ which describes this
effect is displayed in Fig. 2a. In addition, some limiting specific
features of the behavior of $Q_n(\varphi)$ can be revealed with
simple analytical expressions.

\begin{figure}
\setlength{\unitlength}{1cm}
\includegraphics[height=11cm,width=9cm]{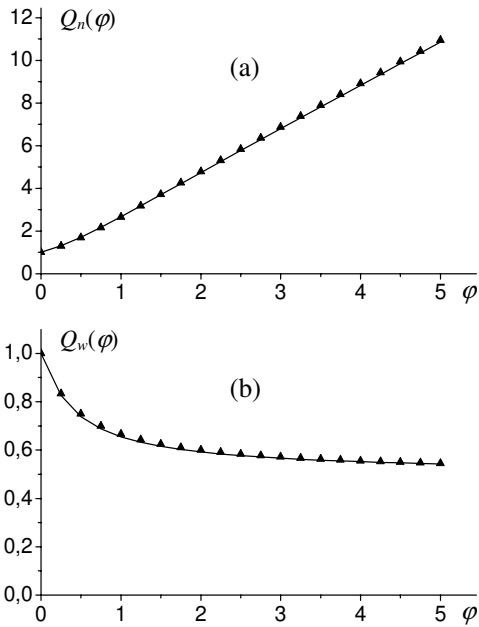}
\caption{The force ($\varphi = fa/2$) dependence of dimensionless
escape rates $Q_{\nu}(\varphi ) = w_e
(\varphi)/w^{0}_{e_{\nu}}(\varphi)$ ($\nu = n, w$) [for narrow ($n$)
and wide ($w$) wells], calculated with exact eqs. (\ref{narw35}) and
(\ref{narw45}) (full lines) and interpolation formulas
(\ref{narw41a}) and (\ref{narw50}) (triangles).}
\end{figure}

{\it a. Weak force limit.}\, In the limit of weak external force,
when $\varphi =fa/2 \ll 1$, in two lowest orders in $\varphi$ the
mean escape rate $w_e$ can be estimated using only the first term
(with $l = 0$) in the sum $S_n(\varphi)$ in the expression
(\ref{narw36}) for $Q(\varphi) = w_e(\varphi)/w_{e_n}^0 $:
\begin{equation} \label{narw40}
Q_n (\varphi) = Q_n^w(\varphi) \approx 1 + \varphi = 1 + fa/2.
\end{equation}

{\it b. Strong force limit.}\, In the opposite case $\varphi =fa/2
\gg 1$, denoted as the strong external force limit, the analysis of
the dependence $w_e (\varphi)$ with the use of eq. (\ref{narw36}) is
more complicated. However, the corresponding limiting dependence can
easily be derived taking into account that in the limit $\varphi \gg
1$ the escaping process is, actually, one-dimensional, for which the
$Q(\varphi)$-factor is given by\cite{shu4}
\begin{equation} \label{narw41}
Q_n (\varphi) = Q_n^s(\varphi)  \approx 2\varphi = fa.
\end{equation}
In deriving eq. (\ref{narw41}) we used the expression for
one-dimensional escape rate\cite{shu4} $w_1 = Df/Z_1$, in which $Z_1
= \int_{d<r<a}dr \,  e^{-u(r)} \approx Z_w/a^2 $ is the
one-dimensional partition function.

This relation can also be illustrated by the semiquantitative
estimation with the use of eq. (\ref{narw36}) by truncating the sum
at $l = l^{*} \sim \sqrt{\varphi} \gg 1$. The fact is that the terms
of the sum with $l < l^{*}$ give the main contribution to the sum
since the terms with $l > l^{*}$ rapidly decrease with the increase
of $l$. At $1 < l < l^{*}$ we can use the estimations \cite{Abr}
$I_{l-\frac{1}{2}}(\varphi)\sim e^{\varphi}/\sqrt{\phi}$ and
$K_{l-\frac{1}{2}} (\varphi)/K_{l+\frac{1}{2}}(\varphi) \approx 1$
so that $q_{l<l^{*}}(\varphi) \sim \varphi$.  Substitution of these
relations into eq. (\ref{narw36}) yields the relation
$S(\varphi)\sim [\varphi Z_{0}(\varphi)]^{-1}
\sum_{l=0}^{l^{*}}(l+\mbox{$\frac{1}{2}$})
I_{l+\frac{1}{2}}^2(\varphi) q_l(\varphi) \sim \varphi$ which is in
agreement with the estimation (\ref{narw41}) [taking into account
that at $\varphi \gg 1$ in eq. (\ref{narw35})
$\varphi\coth(2\varphi) \approx \varphi$].


{\it c. Interpolation formula.} The limiting results obtained above
for $\varphi  \ll 1$ and $\varphi \gg 1$ can be combined into a
simple algebraic interpolation formula
\begin{equation} \label{narw41a}
Q_n(\varphi) \approx Q_n^i(\varphi) = \frac{1}{2}+ \frac{\varphi
e^{2\varphi}}{\sinh(2\varphi)}+ \frac{\varphi^2}{6+2\varphi^2},
\end{equation}
which reproduces function $Q_n (\varphi)$, numerically evaluated
using eqs. (\ref{narw35})-(\ref{narw39}), with accuracy $\sim 3\%$
(see Fig. 2a).

\subsection{Wide-well shape}

Another form of the well shape, in which analysis of the escape rate
$w_e$ can be made analytically, corresponds to the small distance of
closest approach, or large $a$, for which $d \sim r_b \ll a$. In
this case in a fairly wide region of relatively strong force $f <
1/d,1/r_b $ the escape kinetics is fairly accurately described by
the two-state model (\ref{prf7b}).

It is important to note that the inequality $fr_b < 1$ ensures quite
high accuracy of the approximation neglecting the effect of force on
the well shape in the region near the bottom. In this approximation,
the quasiequilibrium population distribution within the well is
isotropic:
\begin{equation} \label{narw42}
|\Psi_e\rangle = |Y_0\rangle.
\end{equation}
This, in turn, means that the partition function $Z_w$ is
independent of the angle $\theta$ and is given by eq. (\ref{nof7}).

The effect of force, however, manifests itself in the anisotropy of
the activation energy of escaping $u_a (\theta)$:
\begin{equation} \label{narw43}
u_a (\theta) \approx u_{\bf f} (\theta, a) \approx u_b + 2\varphi
\cos \theta , \;\;\;(\varphi  = fa/2),
\end{equation}
which, in turn, leads to the anisotropy of the detailed balance
relation, i.e. the anisotropy of the equilibrium constant
\begin{equation} \label{narw44}
K_e (\theta) = K_e^0 e^{-2\varphi\cos\theta},
\end{equation}
where $K_e^0$ is the isotropic equilibrium constant in the absence
of external force given by eq. (\ref{narw43}).

Formula (\ref{narw44}) calls for some additional comments especially
concerning its applicability. The fact is that the value of $K_e
(\theta)$ at each particular $\theta$ is determined assuming local
quasiequilibrium of the population outside and inside the well in
the region close to $r = a$ at this $\theta$. In general, it is
difficult to justify the existence of the quasiequilibrium in the
considered limit, unlike the limit of narrow well discussed above.
This is because for $\varphi = fa/2 \lesssim 1$ the time of passing
over the escaping barrier width  $\delta_b \sim {\rm min}\{a,
f^{-1}\} $ (the width of the region of transition from the inner
part of the well to the outer one), $\tau_{b} \sim \delta_b^2/D$ is
comparable with the time of reorientation $\tau_c \sim a^2/D$. It is
worth noting, however, that the accuracy of quasiequilibrium
assumption becomes better with increasing $f$ since the for $\varphi
= fa/2 \gg 1$ the width $\delta_b \ll a$ and, correspondingly,
$\tau_{b} \ll \tau_c$.

The above-mentioned arguments lead us to the conclusion that in the
considered limit of small radius of the well bottom the two-state
model with $\theta$-dependent equilibrium constant $K_e (\theta)$
gives quite reasonable interpolation formula for the kinetics of the
escaping process and, in particular, for the escape rate $w_e$,
which correctly describes both the limit of weak and strong external
force. Further analysis (see below) will confirm this statement.

Formula for the escaping kinetics can straightforwardly be derived
with the use of general formulas (\ref{narw30}), (\ref{narw30a}),
and some results obtained above in the limit of narrow potential
well. The fact is that, in the mathematical form, the average of any
operator multiplied by angular dependent equilibrium constant
(\ref{narw44}) [of type of eq. (\ref{narw30b})] over the isotropic
equilibrium state (\ref{narw42}) is similar to the average over the
equilibrium distribution (\ref{narw34}), except for the partition
function $Z_0 (\varphi)$ [eq. (\ref{narw34a})], which should be
replaced by $Z_0 (\varphi\to 0) = 2$ corresponding to the isotropic
distribution. These simple algebraic manipulations result in the
following expression for $w_e (\varphi)$
\begin{equation} \label{narw45}
w_e (\varphi)/w^0_{e_{w}}(\varphi) = Q_{w}(\varphi) =
\mbox{$\frac{1}{2}$}Z_0(\varphi)e^{-2\varphi} Q_{n}(\varphi)
\end{equation}
where $Z_0 (\varphi)$ and $Q_{n}(\varphi)$ are determined in eqs.
(\ref{narw34a}) and (\ref{narw35}), respectively, and
\begin{equation} \label{narw47}
w^0_{e_w}(\varphi) = w_e^0 e^{2\varphi}
\end{equation}
is the escape rate in the absence of the external force but with the
activation energy $u_a^{*}$, corresponding to the orientation
$\theta = \pi$ (most favorable for escaping):
\begin{equation} \label{narw47a}
u_a^{*} \equiv  u_a (\theta = \pi)   = u_b - 2\varphi.
\end{equation}

As in the case of narrow potential well the dependence of $w_e$ on
the force $f$ is expressed in terms of that on the only parameter
$\varphi$. The characteristic function $Q_{w}(\varphi)$, which
determines the pre-exponential factor in the activation type
dependence of $w_e (\varphi)$, is displayed in Fig. 2b.  The
numerical results show that $Q_w (\varphi)$ monotonically decreases
(with increasing $\varphi$) from $Q_w  = 1$ at $\varphi = 0$ to $Q_w
= 1/2$ at $\varphi \to \infty$. This behavior is markedly different
from that of $Q_n (\varphi)$ although the specific features of $Q_w
(\varphi) $-dependence are essentially determined by those of $Q_n
(\varphi)$. Some of features of the function $Q_w (\varphi) $, for
example saturation at $\varphi \to \infty$, looking unexpected at
first sight, can be understood by simple analysis (see below).

{\it a. Weak force limit.}\, In the weak force limit the behavior of
$Q_w (\varphi)$ at $\varphi =fa/2 \ll 1$ differs form that obtained
above for $Q_n (\varphi)$ (i.e. for narrow potential well):
$Q_w(\varphi)$ decreases with increasing $\varphi$, so that at
$\varphi \ll 1$
\begin{equation} \label{narw48}
 Q_w^w(\varphi) =  Q_w (\varphi\ll 1) \approx 1 -  \varphi.
\end{equation}
Such a behavior of $Q_w(\varphi)$  results from using the
$\varphi$-dependent normalizing rate $w^0_{e_w} \sim e^{2\varphi}$
(instead of $w_{e_n}^{0} = w_e^{0}$) in the definition of
$Q_w(\varphi)$.

{\it b. Strong force limit.}\,  In the opposite limit $\varphi =fa/2
\gg 1$ the force strongly affects the average escape rate $w_e$,
first of all, because it significantly changes the activation energy
of the rate $w_e$. As for $Q_w (\varphi)$, which characterizes the
pre-exponential factor of the corresponding Arrenius-type expression
for $w_e$, at $\varphi \gg 1$ it monotonically decreases approaching
the asymptotic value $1/2$.

The obtained $Q_w (\varphi) $-independence at $\varphi \to \infty$
can easily be understood by taking into account that, according to
formula (\ref{narw44}), in the case of wide well the strong external
force causes significant localization of the flux of escaping
particles in a small region of orientations $\delta \theta = \pi -
\theta \lesssim 1/\sqrt{\varphi}\ll 1$. The escape rate is
determined by the total flux $J_e$ through this region of size $s_e
\sim (\delta\theta)^2 \sim \varphi^{-1}$. In the strong force limit
$\varphi \gg 1$ the flux $J_e \sim \varphi$, as it follows from eq.
(\ref{narw41}), so that $Q_w (\varphi) \sim s_e J_e \sim {\rm
const}$. The exact estimation can be obtained just by substitution
of the corresponding limiting expression (\ref{narw41}) into eq.
(\ref{narw45}):
\begin{equation} \label{narw49}
Q_w^s (\varphi)=\mbox{$\frac{1}{2}$}Z_0(\varphi)e^{-2\varphi}
Q_{n}^s(\varphi)|_{\varphi\gg 1} = \mbox{$\frac{1}{2}$}
\end{equation}

Similarly to the narrow well limit, in the case of wide well for
large $\varphi$ the escape rate is determined by the
quasi-one-dimensional flux of escaping particles. The mechanism of
formation of the one-dimensional flux is, however, somewhat
different in both cases: for narrow wells the transition to the one
dimensional regime results from high localization of the well
population in the small region at $\theta \sim \pi$, while for wide
wells this transition is caused by strong localization of favorable
transition rates in this region.

{\it c. Interpolation formula.} A simple interpolation expression
for $Q_w (\varphi)$ can be derived, for example, with the use of
similar formula for $Q_n (\varphi) $ presented in eq.
(\ref{narw41a}):
\begin{equation} \label{narw50}
Q_w (\varphi)\approx Q_w^{i}
(\varphi)=\mbox{$\frac{1}{2}$}Z_0(\varphi)e^{-2\varphi}
Q_{n}^i(\varphi).
\end{equation}
Quite satisfactory accuracy ($\sim 3\%$) of this formula is
demonstrated in Fig. 2b.

\section{Summary and concluding remarks}

This work concerns detailed theoretical study of the effect of the
external force $f = F/(k_B T)$ on the kinetics of depopulation of a
deep potential well. The well is assumed to be isotropic and highly
localized (short range). Though, detailed analysis
shows\cite{shu2,shu3} that fairly deep potential well resulted, for
example, from the Coulomb interaction (at large $r$) can also be
considered as highly localized in some conditions.

Fairly simple matrix expression for the kinetics of the well
depopulation is obtained, which predicts strong effect of external
force on the kinetics. Moreover, in general, the kinetics is
predicted to be non-exponential.

In our work we have concentrated on the analysis in the most
physically reasonable limit of fast orientational relaxation of the
PDF in the well. In this limit the analytical expression for the
depopulation kinetics is derived which predicts the kinetics to be
close to the exponential and the total depopulation rate to be a sum
of reaction and escape rates. In our work we have mainly studied the
specific features of the escape rate $w_e$ whose value appears to
significantly depend on shape of the well. Simple analytical
expressions for $w_e (f)$ are obtained for two limiting types of
wells: narrow wells of type of well layer at a distance of closest
approach $d$ (for which $a-d \ll a$) and deep wells with small
distance $d \ll a$.

In the case of narrow well the effect of the force on the escape
rate is fairly strong but shows itself only in the preexponential
factor of the Arrenius-type dependence of the rate, i.e. no strong
effect on the activation energy is predicted. On the contrary, in
the case of wide well ($d \ll a$) the force affects not only
preexponential factor but the activation energy as well.

It is worth noting that the effect of an external force on the
diffusion-assisted processes in the presence of interaction between
particles are studied in a number of works (see, for example, refs.
[1] and [7]). Especially comprehensively the force effect (electric
field effect) is analyzed in the case of ion pair recombination
reaction, i.e. in the case of the Coulomb interaction between
particles.

Unfortunately it is practically impossible to compare the results
our analysis with those obtained earlier, since the earlier works
mainly concerned with processes in potentials without well at short
distances, the reactivity is usually assumed to be high. In
particular, in the case of ion pair recombination processes the
recombination kinetics is considered to be determined by diffusive
motion in the pure Coulomb potential.\cite{Ri,Noo,Noo1} It is,
nevertheless, interesting to note that in the small field limit $fa
\ll 1$ the force effect on the probability $P_e (f)$ of escape from
the Coulomb potential, found in ref. [4], is independent of the
initial distance between ions and is represented in the form $P_e
(f) \approx P_e (f=0)(1 + fa/2)$, which is in apparent agreement
with the field dependence of the escape rate obtained in our works
[see eqs. (\ref{narw40}) and (\ref{narw48})].

Noteworthy is also that in the strong force limit the escaping
process becomes nearly one-dimensional in both cases of well shape
considered. In this limit the escaping rate is determined by the
one-dimensional flux from the small region of favorite orientations
corresponding to $\theta \sim\pi $. This fact allows one to easily
improve the considered two-state  model, in which the effect of the
force on the location of top of the barrier (assumed to be at $r =
a$) is neglected. Moreover, in the strong force limit one can also
take into account the smoothness of the shape of the realistic
barrier near the top, which in the two-state model is actually
assumed to be of cusp shape.

Concerning possible applications of formulas obtained, note that the
most convenient for experimental analysis is not the force dependent
inverse mean lifetime ${\bar w}_0 (\varphi)$ [see eq.
(\ref{narw30a})], but the difference ${\bar w}_0 (\varphi) - {\bar
w}_0 (0)= w_e (\varphi) - w_e^0$, which is independent of the rate
$w_r$ of reaction within the well (assumed to be independent of
$\varphi$). The corresponding dimensionless parameters
\begin{equation} \label{narw51}
\delta Q_{\nu} (\varphi)= [{\bar w}_0 (\varphi) - {\bar w}_0
(0)]/w_{e_{\nu}}^{0} (\varphi), \;\;(\nu = n, w),
\end{equation}
are directly related to $Q_{\nu} (\varphi)$:
\begin{equation} \label{narw52}
\delta Q_{n} (\varphi) =  Q_{n} (\varphi) - 1,\;\; \delta Q_{w}
(\varphi) =  Q_{w} (\varphi) - e^{-2\varphi}.
\end{equation}
The behavior of $\delta Q_{n} (\varphi)$ is, clearly, similar to
that of $Q_{n} (\varphi)$ except for evident displacement along the
ordinate axis. As for $\delta Q_{w} (\varphi)$-dependence, shown in
Fig. 3, its form is essentially different from that of $Q_{w}
(\varphi)$: at $\varphi \to 0$ the function $\delta Q_{w} (\varphi)
\approx \varphi$ is similar to $\delta Q_{n} (\varphi)$, while
$\delta Q_{w} (\varphi \to \infty) = 1/2$. Moreover $\delta Q_{w}
(\varphi)$ has a maximum (though not very pronounced) at $\varphi =
\varphi_m \approx 2.0$.

\begin{figure}
\setlength{\unitlength}{1cm}
\includegraphics[height=7.5cm,width=8cm]{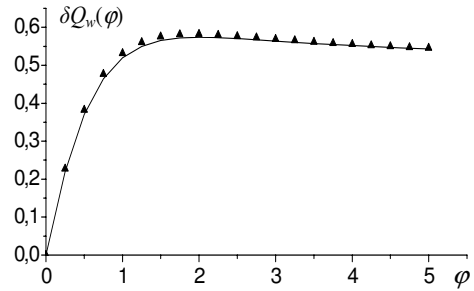}
\caption{The dependence of the dimensionless difference of rates
$\delta Q_{w}(\varphi )$ [see eqs. (\ref{narw51}) and
(\ref{narw52})] calculated with the exact formula (full lines) and
the interpolation expression (\ref{narw50}) (triangles).}
\end{figure}

It is of great interest to compare these theoretical predictions
with experimental results of type of those given in refs. [7-14] but
in the presence of electric field.

Concluding our brief discussion we would like to note that in this
work we restricted ourselves to the analysis of the most realistic
limit of fast orientational relaxation within the well. In reality,
however, with the use of general formula (\ref{narw8a}) one can also
describe the manifestation of finiteness of the orientation
relaxation time. The case, in which the effect of finiteness is
largest, of course, corresponds to  $\hat L_c = 0$, i.e. the absence
of orientational relaxation. In this case the angular dependence of
the equilibrium rate [$K_e (\theta)$], evidently, results in the
highly non-exponential well depopulation kinetics $n_0 (t)$, which
can be approximated by the sum of exponentially decreasing
(monomolecular) contributions with $\theta$-dependent rates, coming
from different orientations. With the use of obtained formulas there
will be no difficulties to analyze this case as well, when needed.

\textbf{Acknowledgements.}\, The author is grateful to Dr. V. P.
Sakun for valuable discussions. The work was supported by the
Russian Foundation for Basic Research.


\end{document}